\documentclass[preprints,article,accept,moreauthors,pdftex]{Definitions/mdpi} 

\firstpage{1} 
\pubvolume{xx}
\issuenum{1}
\articlenumber{5}
\pubyear{2019}
\copyrightyear{2019}
\history{Received: date; Accepted: date; Published: date}

\usepackage{gensymb}

\Title{\textsc{RELXILL\_NK}: A Black Hole Relativistic Reflection Model for Testing General Relativity$^\dagger$}

\Author{Askar B. Abdikamalov $^{1}$, Dimitry Ayzenberg $^{1}$\orcidB{}, Cosimo Bambi $^{1}$\orcidC{}, and Sourabh Nampalliwar~$^{2}$\orcidD{}}

\AuthorNames{Askar B. Abdikamalov, Dimitry Ayzenberg, Cosimo Bambi, and Sourabh Nampalliwar}

\address{%
$^{1}$ \quad Center for Field Theory and Particle Physics and Department of Physics, Fudan University, 200438 Shanghai, China; 17110190057@fudan.edu.cn (A.B.A.), dimitry@fudan.edu.cn (D.A.), bambi@fudan.edu.cn (C.B.)\\
$^{2}$ \quad Theoretical Astrophysics, Eberhard-Karls Universit\"at T\"ubingen, 72076 T\"ubingen, Germany; sourabh.nampalliwar@uni-tuebingen.de (S.N.)}

\firstnote{Presented at the meeting {\it Recent Progress in Relativistic Astrophysics}, 6-8 May 2019 (Shanghai, China).}

\abstract{In this paper, we briefly present \textsc{relxill\_nk}, the first and currently only readily available model of the relativistic reflection spectrum of black hole accretion disks that includes non-Kerr solutions for the black hole spacetime, thus allowing for tests of the Kerr hypothesis and GR. \textsc{relxill\_nk} makes use of a general relativistic ray-tracing code to calculate the relativistic effects of any well-behaved, stationary, axisymmetric, and asymptotically flat black hole spacetime, while the disk physics is handled through the non-relativistic X-ray reflection code \textsc{xillver}. A number of different flavors are available within \textsc{relxill\_nk}; we summarize and compare these flavors using the Johannsen metric for the black hole spacetime.}


\begin{document}
\section{Introduction}

In the next decade, the launch of the next generation of X-ray telescopes, such as ATHENA~\cite{Barret:2018qft} and eXTP~\cite{Zhang:2016}, will provide observations of black hole (BH) accretion disks with significantly higher spectral and temporal resolution than possible with current telescopes. These new telescopes will lead to a deeper understanding of the complex processes involved in accretion and allow for tests of general relativity (GR) that may be comparable to what can be done with gravitational wave observations. Of particular interest is the testing of the Kerr hypothesis~\cite{b1,b2}, i.e.~that the correct description for all isolated, stationary, and axisymmetric astrophysical (uncharged) BHs is the Kerr metric~\cite{1975PhRvL..34..905R, 1967PhRv..164.1776I, 1968CMaPh...8..245I, 1971PhRvL..26.1344H, 1972CMaPh..25..152H, 1971PhRvL..26..331C}. While the Kerr hypothesis holds in GR and some modified gravity theories~\cite{Psaltis:2007cw}, there are some theories in which it does not (e.g.~Chern-Simons gravity~\cite{Alexander:2009tp}). BHs within these theories differ from the Kerr solution and would, in principle, lead to modifications to accretion disk observations as compared to what is expected when Kerr BHs are assumed.

One of the more interesting BH accretion disk observations is known as the X-ray reflection spectroscopy method. Currently the most advanced model for simulating the reflection spectrum is \textsc{relxill}~\cite{doi:10.1093/mnras/sts710, 0004-637X-782-2-76}. \textsc{relxill} assumes the Kerr metric for describing the BH spacetime and while it is possible to test the Kerr hypothesis and GR under this assumption, due to the large amount of physical parameters in the model and the complexity of the accretion disk smaller deviations away from the Kerr solution can be absorbed by other parameters. Thus, it is necessary to have a model for simulating the reflection spectrum in a wide variety of BH spacetimes, such that the Kerr hypothesis can be properly tested and constraints can be placed on modified gravity theories.

In this paper, we briefly present the non-Kerr extension of \textsc{relxill} for the purpose of testing the Kerr hypothesis and GR: \textsc{relxill\_nk}{\footnote{\textsc{relxill\_nk} package available at \href{http://www.physics.fudan.edu.cn/tps/people/bambi/Site/RELXILL_NK.html}{http://www.physics.fudan.edu.cn/tps/people/bambi/Site/ RELXILL\_NK.html} and \href{http://www.tat.physik.uni-tuebingen.de/~nampalliwar/relxill_nk/}{http://www.tat.physik.uni-tuebingen.de/$\sim$nampalliwar/relxill\_nk/}. For support contact \href{mailto:relxill_nk@fudan.edu.cn}{relxill\_nk@fudan.edu.cn}.}}~\cite{b3,Abdikamalov:2019yrr}. This new model allows for the use of any well-behaved, stationary, axisymmetric, and asymptotically flat black hole metric through the use of a general relativistic ray-tracing code that solves for the motion of photons within a given spacetime. \textsc{relxill\_nk} has already been used to analyze the X-ray reflection spectra of a number of BHs and place constraints on some non-Kerr metrics~\cite{noi1, noi2, noi3, noi4, noi5, noi6, noi7, noi8, noi9, noi10, noi11, noi12, noi13}; for a review, see Ref.~\cite{noirev}. In addition to presenting the model we perform a simple analysis of the impact of modifications to the Kerr metric on different flavors of the \textsc{relxill\_nk} model package.

This paper is organized as follows. Section~\ref{sec:mod} briefly describes the model used for calculating the X-ray reflection spectrum. Section~\ref{sec:flav} details the models within the \textsc{relxill\_nk} package. Section~\ref{sec:comp} shows a brief analysis of the effect of a modification to the Kerr spacetime in different models. Section~\ref{sec:conc} concludes by summarizing and discussing possible future improvements to \textsc{relxill\_nk}. Throughout the paper, we use units in which $G_{\rm N}=c=1$ and the convention of the metric with signature $(-+++)$.

\section{Model}
\label{sec:mod}

The BH-disk system is modeled using the standard disk-corona model~\cite{2018AnP...53000430B,book}, in which the BH is surrounded by a geometrically-thin and optically-thick accretion disk and a cloud of hotter gas known as a "corona" is in the vicinity. The disk is in the equatorial plane of the BH and extends from an inner radius $R_{\text{in}}$ to an outer radius $R_{\text{out}}$. The inner radius is usually assumed to be at or near the innermost stable circular orbit (ISCO) radius of the BH. The disk emits locally as a blackbody and when integrated over the radius of the disk becomes a multi-temperature blackbody. The temperature of the disk at any given location depends on the mass of the BH, the accretion rate, and the distance from the BH. At accretion rates of about 10\% of the Eddington rate, the thermal emission of the disk peaks in the soft X-ray band (0.1-1~keV) for stellar-mass BHs and in the optical/UV band (1-10~eV) for supermassive BHs. The corona is a significantly hotter ($\sim100$~keV), usually optically thin, cloud somewhere near the BH and disk. Common geometries include a point or spherical source along the spin axis of the BH or a layer above and below the accretion disk, though the exact morphology is as yet unknown.

\begin{figure}
\begin{center}
\includegraphics[width=8.5cm]{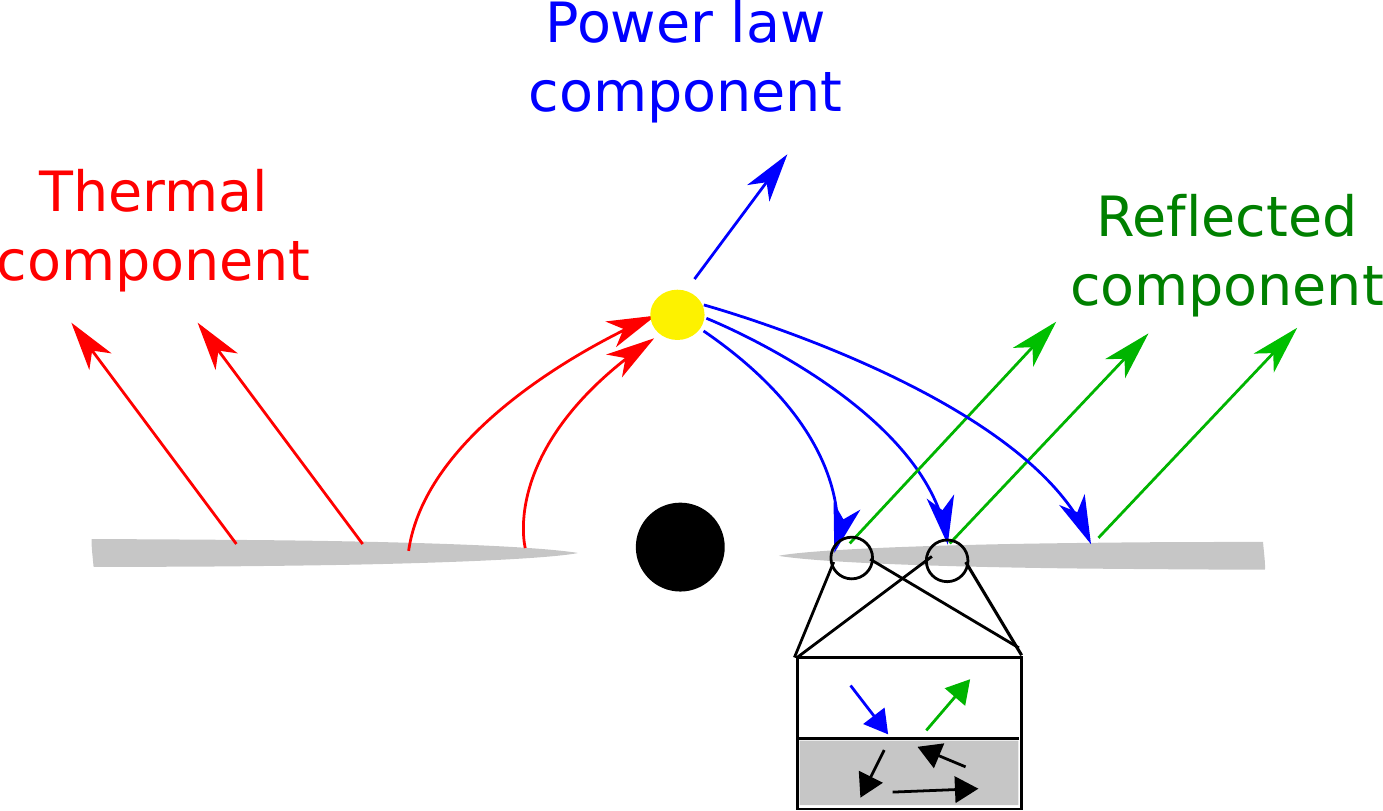}
\end{center}
\vspace{0.3cm}
\caption{\label{fig:model} Sketch of the disk-corona model and the reflection process. From Ref.~\cite{Abdikamalov:2019yrr}.}
\vspace{0.3cm}
\end{figure}

The reflection spectrum is produced through an interaction of the accretion disk and the corona. Some of the thermal photons produced by the disk can inverse Compton scatter off of free electrons in the corona and in turn produce a power-law component that has a cut-off energy ($\sim 300$~keV) dependent on the corona's temperature. Some of this power-law emission can then illuminate the disk, interact internally, and then be re-emitted as a reflection spectrum that includes fluorescent emission lines~\cite{Garcia:2013oma}. A sketch of the disk-corona model and the reflection process is shown in Fig.~\ref{fig:model}. The emission lines are broadened and skewed and the reflection spectrum overall is modified when seen by a distant observer due to the relativistic effects of the BH spacetime (gravitational redshift, Doppler boosting, light bending)~\cite{Fabian:2000nu, Brenneman:2013oba, Reynolds:2013qqa,book}. Thus, observations of the reflection spectrum are a useful tool for studying the properties of BHs with accretion disks.

\textsc{relxill\_nk} is based on \textsc{relxill}, which is a model for the calculation of the reflection spectrum for Kerr BHs~\cite{doi:10.1093/mnras/sts710, 0004-637X-782-2-76}. \textsc{relxill} combines the non-relativistic X-ray reflection code \textsc{xillver}~\cite{2010ApJ...718..695G, Garcia:2013oma} and the relativistic line emission code \textsc{relline}~\cite{2010MNRAS.409.1534D, doi:10.1093/mnras/sts710, Dauser:2014jka}. As \textsc{relxill} is limited to the Kerr spacetime, \textsc{relxill\_nk} was developed for the purpose of studying the reflection spectrum produced by non-Kerr BHs and some significant modifications were required. In particular, the formalism for computing the relativistic effects through the use of the Cunningham transfer functions was kept the same, however the method of computation needed to by modified to be able to allow for a wide range of non-Kerr BH spacetimes. As the Kerr solution admits a third constant of the motion, known as the Carter constant, the equations of motion are separable. In general, however, BH solutions do not have a third constant of the motion and the equations of motion are in turn not separable and do not have a solution by quadrature as in Kerr. Thus, the job of solving for the motions of photons goes from numerically calculating a set of integrals to numerically solving a system of second-order partial differential equations. To do so, \textsc{relxill\_nk} uses a general relativistic ray-tracing code. For more details on the accretion disk model, the Cunningham transfer function method, and the general relativistic ray-tracing code, see the public release paper on \textsc{relxill\_nk}~\cite{Abdikamalov:2019yrr}.

\section{Flavors}
\label{sec:flav}

In the following we briefly describe the available flavors within the \textsc{relxill\_nk} model package and compare the impact of a non-Kerr modification to the spectrum produced by these different flavors. For a discussion of the accuracy of the model as compared with \textsc{relxill} and a more detailed discussion of the impact of a non-Kerr modification on the spectrum from the base models \textsc{relxill\_nk} and \textsc{relxilllp\_nk} see the public release paper~\cite{Abdikamalov:2019yrr}.

The BH reflection spectrum models within the \textsc{relxill\_nk} package depend on a number of physical parameters of the BH and the accretion disk. In the case of a Kerr BH, the only relevant BH parameter is the dimensionless spin of the BH $a^{*}\equiv|\vec J|/M^{2}$, where $\vec J$ is the spin angular momentum of the BH and $M$ is the mass of the BH. The reflection spectrum is independent of the BH mass $M$ and, as stated in Sec.~\ref{sec:mod}, the BH spin angular momentum is aligned with the angular momentum of the disk in the model used here. For non-Kerr BHs, there may be additional parameters that encode modifications away from the Kerr spacetime and/or GR. In this paper, we will analyze the Johannsen spacetime with only one non-zero deformation parameter $\alpha_{13}$. The Johannsen spacetime is detailed in App.~\ref{app:johannsen}. In the model, the deformation parameters are selected through the parameter \texttt{defpar\_type} and the value through \texttt{defpar\_val}.

The majority of the physical parameters in the model describe the accretion disk. The inner edge of the disk $R_{\text{in}}$, the outer edge of the disk $R_{\text{out}}$, and the inclination angle of the disk $\iota$, i.e.~the angle between the observer's line of the sight and the angular momentum of the disk, describe the orientation and geometry of the disk. In the case of supermassive BHs it is necessary to include the cosmological redshift $z$ as it is non-negligible. The atomic physics of the disk is encoded in the iron abundance $A_{\text{Fe}}$ in solar units (in the current \textsc{relxill\_nk} version with \textsc{xillver} all other elemental abundances are set to their solar values), the ionization of the disk $\xi$ ($\xi=4\pi F_{x}/n$, where $F_{x}$ is the flux and $n$ is the gas density), and parameters related to the emissivity profile of the disk. The emissivity profile is dependent on the geometry of the corona and, since that is currently unknown, simplified models are used. For an arbitrary geometry the emissivity profile is modeled with a power-law (the intensity on the disk $I\propto 1/r^{q}$, where $q$ is the emissivity index) or with a broken power-law ($I\propto 1/r^{q_{\text{in}}}$ for $r<R_{\text{br}}$ and $I\propto 1/r^{q_{\text{out}}}$ for $r>R_{\text{br}}$, where $q_{\text{in}}$ and $q_{\text{out}}$ are the inner and outer emissivity indices, respectively, and $R_{\text{br}}$ is the breaking radius). In the lamppost geometry the emissivity profile is instead calculated by assuming a point source corona at some height $h$ along the spin axis of the black hole (see \cite{doi:10.1093/mnras/sts710, Abdikamalov:2019yrr} for more details on the lamppost geometry model). The incident spectrum on the disk is assumed to be a power law with index $\Gamma$ and some models include a reflection fraction $R_{f}$ defined as the ratio of intensity emitted towards the disk from the corona to the intensity escaping to infinity.

Some additional parameters that only appear in a few models include the rest-frame line energy $E_{\rm line}$, a parameter to turn on limb-darkening/-brightening $limb$ (see~\cite{2010MNRAS.409.1534D} for more details), the corona cutoff energy $E_{\rm cut}$, the electron density of the disk $\log N_{\rm e}$, and the electron temperature in the corona $kT_{\rm e}$. The model parameters and which models they are included in are shown in Tab.~\ref{tab:flavors} and the models are briefly summarized in the following:

\begin{itemize}
\item \textsc{relline\_nk}: Base non-Kerr version of relativistic line model \textsc{relline}.
\item \textsc{relconv\_nk}: Similar to \textsc{relline\_nk}, but can convolve any reflection.
\item \textsc{relxill\_nk}: Base non-Kerr version of relativistic reflection model \textsc{relxill}, in which the irradiation of the disk is modeled by a broken power-law emissivity.
\item \textsc{relxillCp\_nk}: Modification of \textsc{relxill\_nk} that uses an nthcomp Comptonization~\cite{1996MNRAS.283..193Z, 1999MNRAS.309..561Z} continuum for the incident spectrum.
\item \textsc{relxillD\_nk}: Same as \textsc{relxill\_nk}, but allows for higher accretion disk electron density (between $10^{15}$ and $10^{19}$ cm$^{-3}$) and the energy cutoff $E_{\text{cut}}=300$ keV.
\item \textsc{rellinelp\_nk}: Modification of \textsc{relline\_nk} in which the incident flux on the disk is due to a isotropically emitting point source at some height along the spin axis of the BH.
\item \textsc{relxilllp\_nk}: Modification of \textsc{relxill\_nk} in which the incident flux on the disk is due to a isotropically emitting point source at some height along the spin axis of the BH.
\item \textsc{relxilllpCp\_nk}: Modification of \textsc{relxilllp\_nk} in which the incident spectrum is an nthcomp Comptonization continuum.
\item \textsc{relxilllpD\_nk}: Same as \textsc{relxilllp\_nk}, but allows for a higher accretion disk electron density (between $10^{15}$ and $10^{19}$ cm$^{-3}$) and the energy cutoff $E_{\text{cut}}=300$ keV.
\end{itemize}
\begin{table*}
\centering
\vspace{0.5cm}
\begin{tabular}{l|cccccccccccccccccccc}
\hline\hline

& \hspace{-0.25cm} $E_{\rm line}$  \hspace{-0.25cm} &  \hspace{-0.25cm} $q_{\rm in}$  \hspace{-0.25cm} &  \hspace{-0.25cm} $q_{\rm out}$  \hspace{-0.25cm} &  \hspace{-0.25cm} $R_{\rm br}$  \hspace{-0.25cm} &  \hspace{-0.25cm} $h$  \hspace{-0.25cm} & \hspace{-0.25cm} $z$ \hspace{-0.25cm} & \hspace{-0.25cm} $\Gamma$ \hspace{-0.25cm} & \hspace{-0.25cm} $\log\xi$ \hspace{-0.25cm} & \hspace{-0.25cm} $A_{\rm Fe}$ \hspace{-0.25cm} & \hspace{-0.25cm} $\log N_{\rm e}$ \hspace{-0.25cm} & \hspace{-0.25cm} $E_{\rm cut}$ \hspace{-0.25cm} & \hspace{-0.25cm} $kT_{\rm e}$ \hspace{-0.25cm} & \hspace{-0.25cm} $limb$ \hspace{-0.25cm} & \hspace{-0.25cm} $R_{\rm f}$ \hspace{-0.25cm}\\

\hline

{\sc relline\_nk} & $\surd$ & $\surd$ & $\surd$ & $\surd$ & $\times$ & $\surd$ & $\times$ & $\times$ & $\times$ & $15$ & $\times$ & $\times$ & $\surd$ & $\times$ \\ 

{\sc relconv\_nk} & $\times$ & $\surd$ & $\surd$ & $\surd$ & $\times$ & $\times$ & $\times$ & $\times$ & $\times$ & $15$ & $\times$ & $\times$ & $\surd$ & $\times$ \\ 

{\sc relxill\_nk} & $\times$ & $\surd$ & $\surd$ & $\surd$ & $\times$ & $\surd$ & $\surd$ & $\surd$ & $\surd$ & $15$ & $\surd$ & $\times$ & $\times$ & $\surd$ \\ 

{\sc relxillCp\_nk} & $\times$ & $\surd$ & $\surd$ & $\surd$ & $\times$ & $\surd$ & $\surd$ & $\surd$ & $\surd$ & $15$ & $\times$ & $\surd$ & $\times$ & $\surd$ \\ 

{\sc relxillD\_nk} & $\times$ & $\surd$ & $\surd$ & $\surd$ & $\times$ & $\surd$ & $\surd$ & $\surd$ & $\surd$ & $\surd$ & $300$ & $\times$ & $\times$ & $\surd$ \\ 

{\sc rellinelp\_nk} & $\surd$ & $\times$ & $\times$ & $\times$ & $\surd$ & $\surd$ & $\surd$ & $\times$ & $\times$ & $15$ & $\times$ & $\times$ & $\surd$ & $\times$ \\ 

{\sc relxilllp\_nk} & $\times$ & $\times$ & $\times$ & $\times$ & $\surd$ & $\surd$ & $\surd$ & $\surd$ & $\surd$ & $15$ & $\surd$ & $\times$ & $\times$ & $\surd$ \\ 

{\sc relxilllpCp\_nk} & $\times$ & $\times$ & $\times$ & $\times$ & $\surd$ & $\surd$ & $\surd$ & $\surd$ & $\surd$ & $15$ & $\times$ & $\surd$ & $\times$ & $\surd$ \\ 

{\sc relxilllpD\_nk} & $\times$ & $\times$ & $\times$ & $\times$ & $\surd$ & $\surd$ & $\surd$ & $\surd$ & $\surd$ & $\surd$ & $300$ & $\times$ & $\times$ & $\surd$ \\ 

\hline\hline
\end{tabular}
\vspace{0.2cm}
\caption{\label{tab:flavors} List of the available models and the parameters of each model. Parameters that are not listed are the dimensionless spin parameter $a^{*}$, the inclination angle $\iota$, the inner disk radius $R_{\text{in}}$, the outer disk radius $R_{\text{out}}$, the deformation parameter selector \texttt{defpar\_type}, and the deformation parameter value \texttt{defpar\_val}, as they are included in all models. $\surd$ means the parameter is part of the model and $\times$ means it is not. }
\end{table*}
%

\section{Flavor Comparison}
\label{sec:comp}

Here we will compare the impact of using a non-Kerr spacetime on the reflection spectrum in different models. Specifically, we will look at the models \textsc{relxill\_nk}, \textsc{relxillCp\_nk}, and \textsc{relxillD\_nk}, as these three models have some different physical assumptions about the accretion disk and corona. (We will not compare the lamppost geometry model to the base model as this was done in the public release paper~\cite{Abdikamalov:2019yrr}.) Through this comparison we wish to study whether any model is more or less affected by a modification to the Kerr spacetime and, in turn, whether any model is more suited for testing the Kerr hypothesis and GR. In principle, if these models are representative of different types of accretion disks in nature, this small analysis can tell us which of these types of accretion disks are more suited for such tests.

\begin{figure*}
\begin{center}
\includegraphics[]{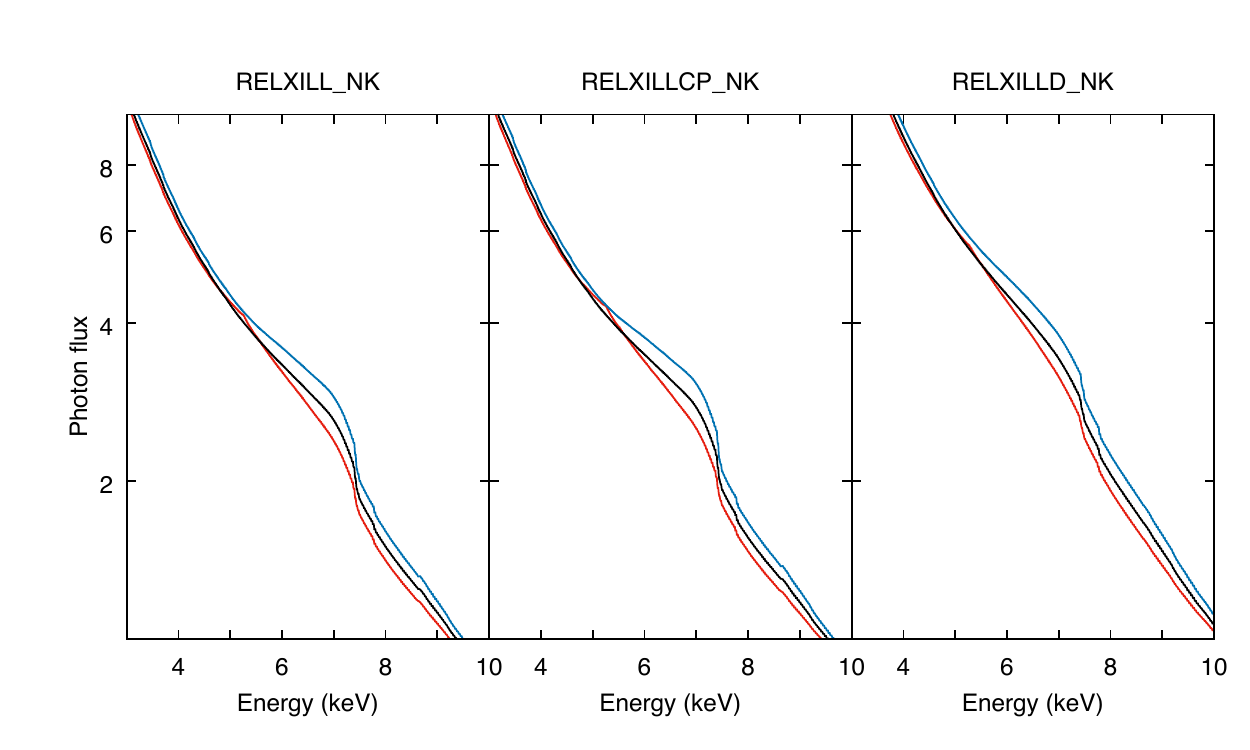}
\end{center}
\vspace{-0.3cm}
\caption{\label{fig:comp} Comparison of \textsc{relxill\_nk}, \textsc{relxillCp\_nk}, and \textsc{relxillD\_nk}, in the Johannsen spacetime with $\alpha_{13}=[-1$(red), $0$(black), $1$(blue)]. Other model parameters are shown in Table~\ref{tab:params}.}
\end{figure*}

As we wish to focus on the differences between the three models \textsc{relxill\_nk}, \textsc{relxillCp\_nk}, and \textsc{relxillD\_nk}, we will set most of the physical parameters to the same values. For simplicity, we only study one value of spin $a^{*}=0.98$ and one value of inclination angle $\iota=45\degree$, however we vary the deformation parameter $\alpha_{13}=[-1,0,1]$, where $\alpha_{13}=0$ represents the Kerr solution. As the primary feature of \textsc{relxillD\_nk} is the inclusion of higher values of accretion disk electron density, we choose the highest available $\log N_{\rm e}=19$ to distinguish it from the forced value of $\log N_{\rm e}=15$ used in \textsc{relxill\_nk} and \textsc{relxillCp\_nk}. In the case of \textsc{relxillCp\_nk}, the shape of the incident spectrum is modified from that of the power-law used in \textsc{relxill\_nk} and we want to focus on that in particular. Thus, we set the high-energy cutoff $E_{\rm cut}=300$~keV and electron temperature $kT_{\rm e}=300$~keV as the value for $E_{\rm cut}$ is forced to 300~keV in \textsc{relxillD\_nk}. The other physical parameters are set to the same values and all relevant parameter values are shown in Tab.~\ref{tab:params}. The simulated reflection spectra zoomed in on the iron line region are shown in Fig.~\ref{fig:comp}.

\begin{table*}
\centering
\vspace{0.5cm}
\begin{tabular}{l|ccccccccccccccc}
\hline\hline

& $a^{*}$ & $\iota$ & $\alpha_{13}$ & $q$ & $R_{\rm in}$ & $R_{\rm out}$ & $z$ & $\Gamma$ & $\log\xi$ & $A_{\rm Fe}$ & $\log N_{\rm e}$ & $E_{\rm cut}$ & $kT_{\rm e}$ & $R_{f}$\\

\hline

{\sc relxill\_nk} & 0.98 & 45 & [-1,0,1] & 3 & $-1$ & 400 & 0 & 2 & 3.1 & 1 & 15 & 300 & -- & $-1$ \\ 

{\sc relxillCp\_nk} & 0.98 & 45 & [-1,0,1] & 3 & $-1$ & 400 & 0 & 2 & 3.1 & 1 & 15 & -- & $300$ & $-1$ \\ 

{\sc relxillD\_nk} & 0.98 & 45 & [-1,0,1] & 3 & $-1$ & 400 & 0 & 2 & 3.1 & 1 & 19 & 300 & -- & $-1$ \\ 

\hline\hline
\end{tabular}
\vspace{0.2cm}
\caption{\label{tab:params} Model parameters used for Fig.~\ref{fig:comp}. $R_{in}=-1$ corresponds to setting $R_{in}$ to the ISCO radius and $R_f=-1$ corresponds to only including the reflected component of the spectrum.}
\end{table*}

As expected the reflection spectra near the iron line region for \textsc{relxill\_nk} and \textsc{relxillCp\_nk} are indistinguishable. This is because the nthcomp Comptonization continuum used for the incident spectrum in the latter only significantly modifies the low and high energy behavior from that in the former. Additionally, the relativistic effects that are necessary to distinguish between different BH metrics have a weaker impact on the low and high energy regions and are generally only dominant in the iron line region around $6.4$ keV. Thus, an accretion disk for which \textsc{relxillCp\_nk} model is more accurate would not be particularly better for the purpose of testing the Kerr hypothesis and GR than a disk for which the standard \textsc{relxill\_nk} model applies.

In the case of the \textsc{relxillD\_nk} model we do find some differences from the base \textsc{relxill\_nk} model. The affect of non-Kerr modifications are slightly weaker in the iron line region in the \textsc{relxillD\_nk} spectrum. In both models the modification from Kerr becomes smaller away from the iron line region; however, this reduction seems to be slower in the case of \textsc{relxillD\_nk}. It is not clear from such a brief analysis if these differences would lead to a better or worse test of the Kerr hypothesis, but it does make the case for a more thorough analysis in a future work.

\section{Concluding Remarks}
\label{sec:conc}

We have briefly presented the new non-Kerr X-ray reflection model \textsc{relxill\_nk}. This new model allows for better testing of the Kerr hypothesis and GR through observations of the X-ray reflection spectrum of BHs with accretion disks. In addition to presenting the model, we have also performed a simple analysis of whether different flavors of the \textsc{relxill\_nk} model are better suited for testing the Kerr hypothesis. In general, we have found that the simple analysis performed here is not enough to answer this question outright, however it is clear that there are differences in how modifications to the Kerr metric influence the spectrum in different flavors. Thus, a more thorough and complex analysis would be worthwhile, but is beyond the scope of this paper.

There are some improvements that are currently being worked on for the \textsc{relxill\_nk} model. There are concerns about the accuracy of the model at high inclinations as well as the overall accuracy not being good enough for the next-generation telescopes. A more accurate model is in the works and is expected to be completed soon. The infinitesimally thin accretion disk model used in \textsc{relxill\_nk} is not suitable for all BH accretion disks and so there is work in progress to study the impact of slim and thick accretion disks as well as adding the former into the model package. Another assumption in the accretion disk model is that the inner edge of the disk is at the ISCO radius. Whether this is the case in reality is as of yet unclear and so we are also studying the impact of extending the disk below the ISCO radius and may incorporate such a model into the package. Finally, we currently only consider photons that travel from the corona directly to the disk and then on to the observer without crossing the mid-plane of the BH. There are photons that can originate from the underside of the mid-plane (relative to the observer) and so we are studying the impact of these photons on the reflection spectrum. We also plan to add several new flavors into the \textsc{relxill\_nk} package that include the affect of these underside originating photons.

\acknowledgments{This work was supported by the Innovation Program of the Shanghai Municipal Education Commission, Grant No. 2019-01-07-00-07-E00035, and Fudan University, Grant No. IDH1512060. A.B.A. also acknowledges support from the Shanghai Government Scholarship (SGS). S.N. acknowledges support from the Alexander von Humboldt Foundation and the Excellence Initiative at Eberhard-Karls Universit\"at T\"ubingen.}

\appendixtitles{yes} 
\appendix
\section{Johannsen Metric}
\label{app:johannsen}

The non-Kerr metric proposed by Johannsen~\cite{PhysRevD.88.044002} that is a subset of the larger class of metrics first proposed by Vigeland, Yunes, and Stein~\cite{Vigeland:2011ji}, in Boyer-Lindquist coordinates is given by the line element

\begin{align}
ds^{2}=&-\frac{\tilde\Sigma\left(\Delta-a^{2}A_{2}^{2}\sin^{2}\theta\right)}{B^{2}}dt^{2}-\frac{2a\left[\left(r^{2}+a^{2}\right)A_{1}A_{2}-\Delta\right]\tilde\Sigma\sin^{2}\theta}{B^{2}}dtd\phi
\nonumber \\
&+\frac{\tilde\Sigma}{\Delta A_{5}}dr^{2}+\tilde\Sigma d\theta^{2}+\frac{\left[\left(r^{2}+a^{2}\right)^{2}A_{1}^{2}-a^{2}\Delta\sin^{2}\theta\right]\tilde\Sigma\sin^{2}\theta}{B^{2}}d\phi^{2},
\end{align}
where
\begin{align}
&B=\left(r^{2}+a^{2}\right)A_{1}-a^{2}A_{2}\sin^{2}\theta, \quad \tilde\Sigma=\Sigma+f,
\nonumber \\
&\Sigma=r^{2}+a^{2}\cos^{2}\theta, \quad \Delta=r^{2}-2Mr+a^{2},
\end{align}
the four free functions $f$, $A_{1}$, $A_{2}$, and $A_{5}$, are\footnote{The four free functions $f$, $A_{1}$, $A_{2}$, and $A_{5}$, are written as a power series in $M/r$
\begin{align}
&f=\sum_{n=2}^{\infty}\epsilon_{n}\frac{M^{n}}{r^{n-2}}, \quad
A_{1}=1+\sum_{n=0}^{\infty}\alpha_{1n}\left(\frac{M}{r}\right)^{n}, \quad
\nonumber \\
&A_{2}=1+\sum_{n=0}^{\infty}\alpha_{2n}\left(\frac{M}{r}\right)^{n}, \quad
A_{5}=1+\sum_{n=0}^{\infty}\alpha_{5n}\left(\frac{M}{r}\right)^{n}.
\end{align}
In order to correctly recover the asymptotic limit, one must impose $\alpha_{10}=\alpha_{20}=\alpha_{50}=0$. Without loss of generality, we can set $\alpha_{11}=\alpha_{21}=\alpha_{51}=0$ as these can be absorbed into the definition of $M$ and $a$. To satisfy Solar System constraints without fine-tuning, $\epsilon_{2}=\alpha_{12}=0$. Thus, the leading-order deformation parameters that are not tightly constrained by Solar System observations are $\epsilon_{3}$, $\alpha_{13}$, $\alpha_{22}$, and $\alpha_{52}$. See~\cite{PhysRevD.88.044002} for more details.}
\begin{align}
f=&\sum_{n=3}^{\infty}\epsilon_{n}\frac{M^{n}}{r^{n-2}},
\nonumber \\
A_{1}=&1+\sum_{n=3}^{\infty}\alpha_{1n}\left(\frac{M}{r}\right)^{n},
\nonumber \\
A_{2}=&1+\sum_{n=2}^{\infty}\alpha_{2n}\left(\frac{M}{r}\right)^{n},
\nonumber \\
A_{5}=&1+\sum_{n=2}^{\infty}\alpha_{5n}\left(\frac{M}{r}\right)^{n},
\end{align}
and $a=|\vec{J}|/M$ is the spin parameter of the BH.

The Johannsen metric is a parametric deformation of the Kerr spacetime and while it can in principle be mapped to non-Kerr solutions within GR and in modified theories of gravity it is not itself a solution to any theory of gravity. The deformations are encoded in the four free functions $f, A_{1}, A_{2},$ and $A_{3}$, through the deformation parameters $\epsilon_{n}, \alpha_{1n}, \alpha_{2n},$ and $\alpha_{5n}$. When these parameters are set to zero the Johannsen metric reduces to Kerr. In this work we focus on the parameter $\alpha_{13}$ and set all others to zero as this is the parameter that has the largest impact on the spacetime~\cite{PhysRevD.88.044002}.


\reftitle{References}

\end{document}